\documentclass[conference,10pt]{IEEEtran}
\IEEEoverridecommandlockouts
\usepackage{cite}
\usepackage{amsmath,amssymb,amsfonts}
\usepackage{algorithmic}
\usepackage{graphicx}
\usepackage{textcomp}
\usepackage{xcolor}
\usepackage{subcaption}

\def\BibTeX{{\rm B\kern-.05em{\sc i\kern-.025em b}\kern-.08em
	T\kern-.1667em\lower.7ex\hbox{E}\kern-.125emX}}

\usepackage{hyperref}
\usepackage{subcaption}

\DeclareMathOperator*{\argmax}{argmax}

\usepackage{tikz}

\usepackage[OT2,T1]{fontenc}
\usepackage{substitutefont}

\substitutefont{OT2}{\rmdefault}{wncyr}

\DeclareRobustCommand{\rn}[1]{
	{\fontencoding{OT2}\selectfont#1}%
}

\begin{document}
	
	\title{A Cognitive Framework for Delegation Between Error-Prone AI and Human Agents\\
		{}
	}
	
	\author{
		\IEEEauthorblockN{Andrew Fuchs\thanks{This work was supported by the H2020 Humane-AI-Net project (grant \#952026) and by the CHIST-ERA grant CHIST-ERA-19-XAI-010, by MUR (grant No. not yet available), FWF (grant No. I 5205), EPSRC (grant No. EP/V055712/1), NCN (grant No. 2020/02/Y/ST6/00064), ETAg (grant No. SLTAT21096), BNSF (grant No. \rn{KP}-06-\rn{DOO}2/5).}}
		\IEEEauthorblockA{\textit{Department of Computer Science} \\
			\textit{Universit\`{a} di Pisa}\\
			\textit{National Research Council (CNR)}\\
			Pisa, Italy \\
			andrew.fuchs@phd.unipi.it}
		\and
		\IEEEauthorblockN{Andrea Passarella}
		\IEEEauthorblockA{\textit{Institute for Informatics and Telematics}\\ 
		\textit{National Research Council (CNR)}\\
		Pisa, Italy \\
		a.passarella@iit.cnr.it}
	\and
	\IEEEauthorblockN{Marco Conti}
	\IEEEauthorblockA{\textit{Institute for Informatics and Telematics}\\ 
	\textit{National Research Council (CNR)}\\
	Pisa, Italy \\
	marco.conti@iit.cnr.it}
}

\maketitle

\begin{tikzpicture}[remember picture,overlay]
\node[anchor=south,yshift=10pt] at (current page.south) {\fbox{\parbox{\dimexpr\textwidth-\fboxsep-\fboxrule\relax}{
  \footnotesize{
    This work has been submitted to the IEEE for possible publication. Copyright may be transferred without notice, after which this version may no longer be accessible.
  }
}}};
\end{tikzpicture}

\begin{abstract}
With humans interacting with AI-based systems at an increasing rate, it is necessary to ensure the artificial systems are acting in a manner which reflects understanding of the human. In the case of humans and artificial AI agents operating in the same environment, we note the significance of comprehension and response to the actions or capabilities of a human from an agent's perspective, as well as the possibility to delegate decisions either to humans or to agents, depending on who is deemed more suitable at a certain point in time. Such capabilities will ensure an improved responsiveness and utility of the entire human-AI system. To that end, we investigate the use of cognitively inspired models of behavior to predict the behavior of both human and AI agents. The predicted behavior, and associated performance with respect to a certain goal, is used to delegate control between humans and AI agents through the use of an intermediary entity. As we demonstrate, this allows overcoming potential shortcomings of either humans or agents in the pursuit of a goal.
\end{abstract}


\section{Introduction}
\label{sec:intro}

AI algorithms are ever increasingly impacting aspects of our lives. The use of AI to assist in or execute decisions spans many topics and disciplines: algorithmic trading, police dispatching, ride-sharing services, online dating, etc. \cite{rahwan2019machine}. As the ubiquity of AI systems increases, so too are we seeing advancements in their complexity and capability. With humans encountering AI systems at increasing frequencies, there is extensive study and need for AI algorithms which can take the user in mind in order to best perform \cite{cichocki2021future}. Therefore, research in Human-Centric AI (HCAI) is proving useful and essential for effective systems endowed with comprehension or awareness of the human. HCAI and related topics demonstrate methods for training and utilizing AI systems with the human in mind, and methods for improving the collective performance of the human-AI hybrid system.

Despite the importance or desire to augment human reasoning and skills with AI-based methods, there can be pitfalls. In fact, the consequences for mistakes by an AI can be quite severe. For instance, chat bots can start to behave unpleasantly or offensively \cite{neff2016automation, wolf2017we}, self-driving cars can have fatal crashes \cite{dearman2019wild, ferrara2016self}, or algorithms can be fooled into making erroneous decisions \cite{carrara2017detecting, jeddi2020learn2perturb}. Therefore, we can see AI systems are not infallible and consequently still require awareness of these shortcomings. Consequently, with HCAI focusing on different mechanisms supporting the combination of human and artificial intelligence, or the augmentation of human reasoning, it is important to consider potential errors, of either AI, or human, or both. To that end, we investigate detecting sub-optimal behavior based on observations in order to delegate control to the party - either human or AI - which, at any given point in time, is predicted to perform best.

In this paper, we consider the scenario of humans and AI systems operating as a team to accomplish a task. For our scenario, the composition of the team will be such that there is a potentially heterogeneous mixture operating under the supervision of a managing agent tasked with delegation. The manager learns, via observation of behavior causing environment changes, which actor would be ideal to perform the next action in pursuit of the team's goal. We utilize a combination of Q-Learning Reinforcement Learning (RL) agents and Instance-based Learning (IBL) agents using cognitively inspired mechanisms. The use of IBL for agents is meant to allow for a human-like process in both representation and understanding of behavior. We augment the team behavior with injected errors to simulate behavior policies which make mistakes. The method and results demonstrated in this paper are intended to illustrate a cognitively-inspired method of understanding and representing behavior to optimize team dynamics. We show that the performance of the error-induced agents is improved by up to $\approx 80\%$ when combined as a team under the managing agent. Moreover, a manager agent trained by observing the behavior of human and AI agents outperforms a manager agent choosing randomly by up to $38\%$ for key performance indices. The results demonstrate how our method enables strong team performance under a manager which can use cognitively-inspired mechanisms to extract a desirable pattern of team behavior.

\section{Background}

\subsection{Related Work}

HCAI systems are gaining momentum in the recent literature. They are proposed in several areas, and for different applications. Hereafter, we provide a summary of some significant efforts.

\subsubsection{Estimating Mental States and Predicting Behavior} It is important to define systems which can estimate the goals, beliefs, and likely future behavior of others. This predictive power allows agents to act based on an estimated understanding of others, which lets agents utilize their estimates to improve the collective performance of a team \cite{fuchsHanabi, shum2019theory}. This understanding can be accomplished via Theory of Mind (ToM), which enables estimation of the mental state of another \cite{graziano2019attributing}. Accounting for these mental state estimates, decisions can be made. Similarly, behavior can be generated with an integrated behavior prediction model. In this case, the behavior of others can be modeled based on estimates of past behavior or other assumptions. For instance, cognitive models can be used to learn a model representing the observed decision-making process \cite{nguyen2020cognitive, nguyeneffects}.

\subsubsection{AI-assisted Behavior and Decisions}
Commonly, humans and AI interact in the case of AI-assisted decisions and related scenarios. In the case of more sophisticated simultaneous control such as in \emph{Overcooked} \cite{carroll2019utility, wang2020too}, agents in the environment need to understand and/or predict human behavior to optimize their performance. In another context, AI can operate as a backup or alternate in execution of a task. As a result, control is delegated between the two based on thresholds or the human's choice. This allows the human to offload control to the AI system when they deem it necessary or safe to do so \cite{morita2020cognitive}.

The approach presented in this paper falls in this class. With respect to existing literature, we combine the concept of behavior comprehension and AI-assisted decisions to provide improved and learned control delegation based on observed participant performance.

\subsection{Reinforcement Learning: Q-Learning and IBL}

RL is a method by which situations or observations are mapped to actions so as to maximize a reward signal \cite{sutton2018reinforcement}. The maximization is performed by an agent through exploration of an environment and the available actions. The states and actions determine the feedback (reward) an agent observes, which motivates finding best actions. At its base, RL is comprised of several components: agent, environment, policy, reward signal, value function, and an optional model of the environment \cite{sutton2018reinforcement}. These elements are utilized in the learning methods to build a representation of optimal behavior.

\subsubsection{Q-Learning}

In Q-Learning, agents use observations of states $s$, actions $a$, and rewards $r$ to generate an estimate of action values in states and learn a behavior policy. The value estimates are stored and modified using an update such as:
\begin{equation}
\label{eq:q-learning}
	Q_{t+1}(s,a) = (1 - \alpha)Q_t(s,a) + \alpha[r + \gamma max_{a'}Q_t(s',a')]
\end{equation}
where $s'$ represents the new state resulting from $a$ in $s$. Additionally, $\alpha$ is the learning rate used to discount new observations and $\gamma$ is the discount parameter for future state values. The agent selects an action $a=\argmax_a Q(s,a)$ which maximizes $Q$. The typical downside of Q-Learning is that speed of convergence may be low, due to the need of exploring extensively the $(s,a)$ space.

\subsubsection{Instance-Based Learning}

In IBL, agents calculate an estimated utility given the current state based on the blending mechanism of ACT-R. This serves to support a cognitively-inspired approach to mimic the human's learning, behavior, and decision-making. In our case, we utilize the reduced version of the IBL equations seen in \cite{nguyen2020cognitive, nguyeneffects}. Agents observe and store instances $\langle s,a,x\rangle$ in memory with time marker $t$ to represent past experiences. These instances represent the state (or context) $s$, action taken $a$, and outcome $x$ defining the reward or other feedback provided based on $(s,a)$.

To support a model of behavior, the agent uses the past instances matching the current state/context to measure an activation level $A_i$ for each action/reward pair corresponding to the given state. The strength of the activation is based on the observation times, which signify the strength of each memory corresponding to the matching instances. The activation strength of a memory depends both on the recency and the frequency of the memories (i.e. reduced strength as more time elapses). The strength of activation values define a retrieval probability $p_i$ for each instance using a Boltzmann Distribution with temperature parameter $\tau$. The strength of the retrieval probability denotes how strong a memory is and subsequently how much it should impact the measure of utility, which allows the agent to calculate an estimate of value actions
\begin{equation}\label{eqn:est_util}
	V(s,a) = \sum_{i=1}^n p_ix_i
\end{equation}
where $x_i$ is the outcome of instance $i$ (i.e. a reward/value). Given estimated action values, the agent selects the most suitable action $a=\argmax_a V(S,a)$. In our case, we use an additional efficiency scheme regarding the storage of instances. Following \cite{petrov2006computationally}, we approximate $A_i$ to reduce storage complexity while maintaining accuracy.

\section{Human-Centric AI Approach to Decision-Making}
\label{sec:algorithms}

As anticipated in Section~\ref{sec:intro}, we assume the existence of a pool of artificial and human agents, and a manager that decides per state which agent in the pool should make decision at any given point to achieve the team's goal. In the following, we will refer to agents in the pool as \emph{navigating agents}. The pool of navigating agents will be trained first to ensure they have valid policies prior to manager training and testing. We use the Q-Learning algorithm for navigating agents representing the artificial systems. These agents observe the state and then receive feedback as a reward $r$ based on the state-action pair $(s,a)$. For the representation of human-like behavior as well as the manager, we define IBL-based agents. We use IBL for the manager as this agent needs to interpret as close as possible the human behavior (for which IBL is better suited than Q-learning). In the case of the manager, observations represent states and agents selected rather than the movement actions performed by the navigating agents, so the manager will not observe the actions of the agent they choose. This ensures the manager is only able to make its decision based on team performance, not the individual actions of navigating agents.

\subsection{Q-Learning Navigating Agents}

For Q-Learning agents we exploit directly the approach of Equation~\ref{eq:q-learning}. Therefore, agents make observations while taking actions, and the policy is built through the update of the value function. The observations are carried out for a certain number of moves ($L_{max}$) and for a specified number of games ($N_G$). In the case of navigating agents (including Q-learning and IBL agents), the action space $A_G$ is defined by the possible actions allowed by the specific game that agents play (here ``game" denotes generally a given task agents have to accomplishing by performing subsequent decisions - ``moves").

\subsection{IBL Navigating Agents}

Similar to Q-Learning, an IBL agent observes states, actions, and rewards. In this case, as described above, these are stored as instances $i=\langle s,a,r \rangle$ with time marker $t$ for the IBL model. Time markers for new observations of an instance are stored in addition to the first marker $t_0$, which denotes the first time an instance tuple $(s,a,r)$ was observed.  Based on \cite{nguyen2020cognitive, nguyeneffects}, the IBL agents do not observe an immediate feedback signal; instead, all instances in a trajectory observe the same reward based on the game outcome. Consequently, the instances observed from a particular game utilize a single reward based on the final outcome (the specific form of the reward function is problem-dependent). Further, an IBL agent performs moves according to the policy available at the current time. It saves all observed instances with their corresponding time until the conclusion of the game. Hence, action values are not updated until the end of a game.

\subsection{Manager Agent}

The manager selects which agent will choose the action in the current state. Similar to the IBL navigating agent, the manager learns by observing instances $\langle s,a,r \rangle$ and time $t$. In this case, the action space is $A_M=\{1,\dots,N\}$, where $N$ refers to the number of navigating agents. As with the IBL navigating agents, the reward $r$ observed in an instance is problem-dependent, but in general it must follow that the observed value of all actions in a trajectory is the game result from the entire trajectory. Overall, the algorithm for training the manager agent follows the one used to train the IBL navigating agents. The primary difference is to select which agent should decide the action, then let the identified agent select the action.

\section{Experimental settings}
\label{sec:experimental settings}

In this section we describe a concrete experimental environment where we have tested the general framework presented in Section~\ref{sec:algorithms}. As a specific case, we have considered Gridworld, a game where a player needs to navigate in a grid, from a starting cell to a goal cell. Gridworld environments serve as a simple and powerful tool for testing and analyzing policy learning methods. The following sections outline the training and testing environments utilized as well as the agent configurations.

\subsection{Gridworld Environments}

\begin{figure}[t]
	\centering
	\includegraphics[width=0.37\textwidth]{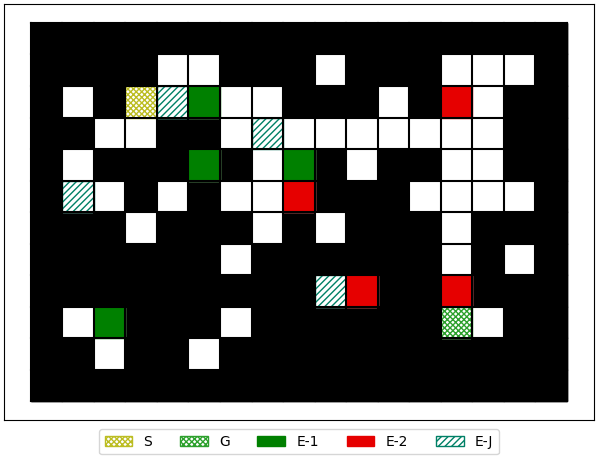}
	\caption{Sample Gridworld environment with error states. (S: start, G: goal, E-1: error $agent_1$, E-2: error $agent_2$, E-J: error $agent_1$ and $agent_2$)}
	\label{fig:basic_error_grid}
	\vspace{-6mm}
\end{figure}

In our experiments, we generated Gridworld environments for the agents to navigate, parameterized by the following: grid dimensions, relative position of start and goal states, ratio of open cells to walls, number of error states of each type (if any). This generates Gridworld environments with the following characteristics. First, in addition to $s_0$ and $s_g$, we include error states $s_{e_i}$ (where $e_i$ indicates an error state for agent $i$). Additionally, we include joint error states in which multiple agents could make a mistake (e.g. $s_{e_{ij}}$). In this case, all agents indicated for that state (i.e. agents $i$ and $j$) may choose an error action. In error states, agents \emph{may} perform an error  by ignoring their policies and instead selecting an action which takes them off an optimal path to the goal. When an agent is selected in their error state, they follow their policy or select an error based on their probability of error $p_{e_i}\in [0,1]$. For each agent, $p_{e_i}$ determines how likely they are to make an error, which provides stochasticity. In an error state assigned to another agent, the state is treated as a normal empty grid cell (i.e $p_{e_i} = 0$ in $s_{e_j}$). Finally, we utilize a single start state and corresponding goal state. Further, some states may be unreachable, but there must be a path from $s_0$ to $s_g$ utilizing actions defined in $A_G$, see \autoref{fig:basic_error_grid}.

\subsection{Agents}

In our experiments, the goal is for the navigating agents is to find $s_g$ as a team. The manager is required to select the agent at each step that will choose the action for the current state. Thus, a state denotes the team is in a certain cell, and the manager action is which navigating agent should decide the next move. As such, the manager is attempting to navigate the team through the Gridworld optimally via the selection of appropriate acting agents. For feedback, the IBL agents (manager and navigating) are only provided with a simple reward based on the outcome of the game:
\begin{equation}\label{eqn:ibl_reward}
	R(f)=\begin{cases}
		100 - L_g \quad &\, \text{f == True} \\
		-100 - L_g \quad &\, \text{otherwise} \\
	\end{cases}
\end{equation}
where $L_g$ is the number of steps needed to reach the goal state. This reward motivates the agents to make good selections and minimizes the amount of direct feedback. Further, this ensures the manager is only able to make assessments based on the navigating agents selected. The goal being to force the manager to rely on outcomes of the entire trajectory rather than receiving an immediate reward for a specific action. A key factor of the manager's success then comes down to its ability to determine the sequence of agents resulting in the best trajectory. Consequently, the manager learns to delegate control effectively to account for error-prone agents or impactful error states as choosing error-prone agents in their error states would be suboptimal.

To train the navigating agents, we utilize the Gridworld action space $A_G$ to navigate the grid cells and transition between states. For Q-Learning rewards, we use the following:
\begin{equation}\label{eqn:q_rewards}
	R_Q=\begin{cases}
		100 \quad &\, \text{goal is reached} \\
		-1 \quad &\, \textrm{open cell to open cell}\\
		-10 \quad &\, \textrm{collision with wall}\\
		-100 \quad &\, \text{otherwise} \\
	\end{cases}
\end{equation}
The use of the $-1$ penalty serves to enforce a priority for shorter paths, and the $-10$ penalty promotes wall avoidance.

An additional aspect of our experiments is the mixture of agents utilized. We allow for both homogeneous and heterogeneous mixtures of policy types for the navigating agents (Q-Learning or IBL). Further, the error probabilities for error states are defined per agent. This allows us to test the manager with differing team compositions.

Two additional scenarios were utilized to test the efficacy of our method. First, navigating agents tested operating in the environment without teammates or a manager agent. This demonstrates the effect of errors on solo agent performance. Second, we replace the IBL manager with a random manager. In this case, the policy of the manager selects navigating agents uniformly randomly in all states. This demonstrates the difference in impact between randomly generated teaming and a learned model of teaming.

We have replicated simulations with $25$ grids at increasing levels of complexity for each grid (approximately 13 levels per grid), and in the following we show average results plotted with error regions signifying the variance of the results.

\section{Results}
\label{sec:results}

Our scenario was tested in the Gridworld setting with varying levels of complexity. We created $25$ error-free Gridworld environments. These base Gridworlds were used to train the navigation agents. With trained navigating agents, we then trained and measured the performance of the managing agents in the same $25$ Gridworld environments, but with increasing numbers of error states. In each case, error states were created incrementally by randomly placing error states in currently open cells. The results were then averaged across the different grids at each error state frequency level.

\subsection{Gridworlds}

Our Gridworlds were created with $10$ rows and $15$ columns, excluding the boundary walls. The starting state $s_0$ and goal state $s_g$ were placed in the same cells for each grid, but the walls were randomly placed. This ensured consistent positioning of the start and goal while requiring varying paths for successful navigation. To vary the difficulty, error states are introduced into the grid and placed randomly. For our experiments, error state frequency was equal for all states, including joint error states \autoref{fig:basic_error_grid}. At the end of each training and testing phase, an additional error state of each type was added at random to an open cell. The ratio of error states to open cells increased from $12.5\%$ to $87.5\%$, which ensured the existence of error states while also allowing for the potential for paths with some error-free cells.

\subsection{Agents}

Following the IBL literature and inspired by \cite{nguyen2020cognitive, nguyeneffects}, the IBL-based agents used the following parameters: $\{d:0.5, \sigma:0.25, \tau:\sigma\sqrt{2}, k:5\}$ For the Q-Learning agents, we utilized: $\{\alpha:0.9999, \alpha \textrm{-decay}:0.9999, \epsilon:0.9999, \epsilon \textrm{-decay}:0.9999, \gamma:0.9\}$. The training of the navigating agents was performed on each Gridworld for $150,000$ episodes at a maximum game length $L_{max}=150$. Manager agents utilized the same parameters and $L_{max}$ as the IBL-based navigating agents, but managers were trained for $20,000$ games.

\subsection{Team Configuration and Results}

For our tests, we utilized the previously defined Gridworld configuration with a wall ratio of $60\%$ (similar results were seen for $40\%$, $50\%$, and $70\%$, but omitted for space). To track performance, we measured mean game length determined by the length of path used to reach the goal state as well as frequency of agent selection by the manager agents (random vs. IBL). These measurements demonstrated the effect of grid complexity on team performance. The main factor in grid complexity in this case is the frequency of error states, which generates error state counts increasing from $2$ to $14$ for each type being incrementally placed in the grid. Further, the measurements also denote the demonstrated learned preferences regarding agent selection of the manager.

\begin{figure}[t]
\centering
\begin{subfigure}{.42\textwidth}
	\centering
	\includegraphics[width=0.9\textwidth]{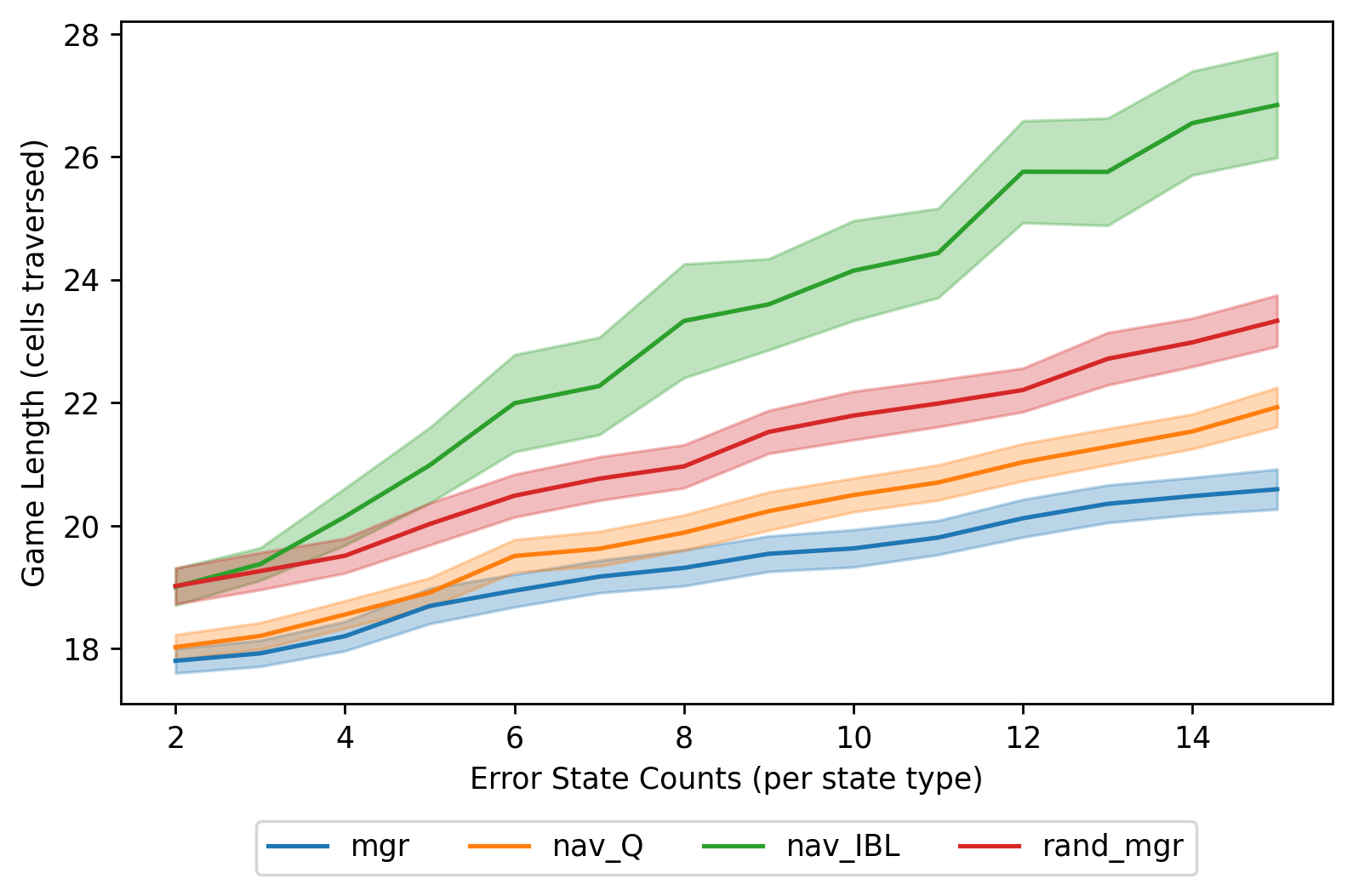}
	\caption{Game Lengths: $\{p_{e_1}=0.25, p_{e_2}=0.25\}$}
	\label{fig:game_lens_60_25_25}
\end{subfigure}
\begin{subfigure}{.42\textwidth}
	\centering
	\includegraphics[width=0.9\textwidth]{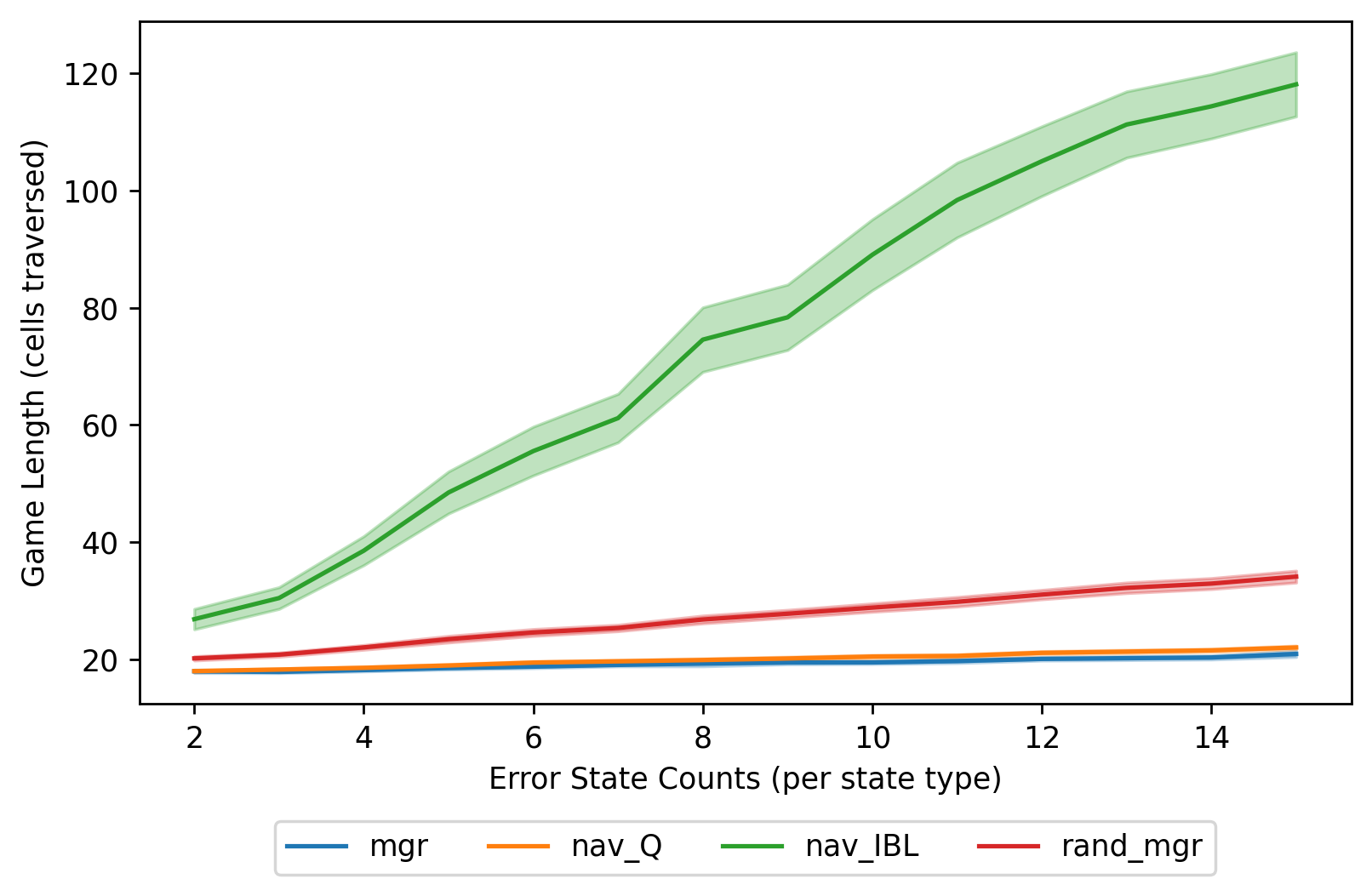}
	\caption{Game Lengths: $\{p_{e_1}=0.25, p_{e_2}=0.75\}$}
	\label{fig:game_lens_60_25_75}
\end{subfigure}
\begin{subfigure}{.42\textwidth}
	\centering
	\includegraphics[width=0.9\textwidth]{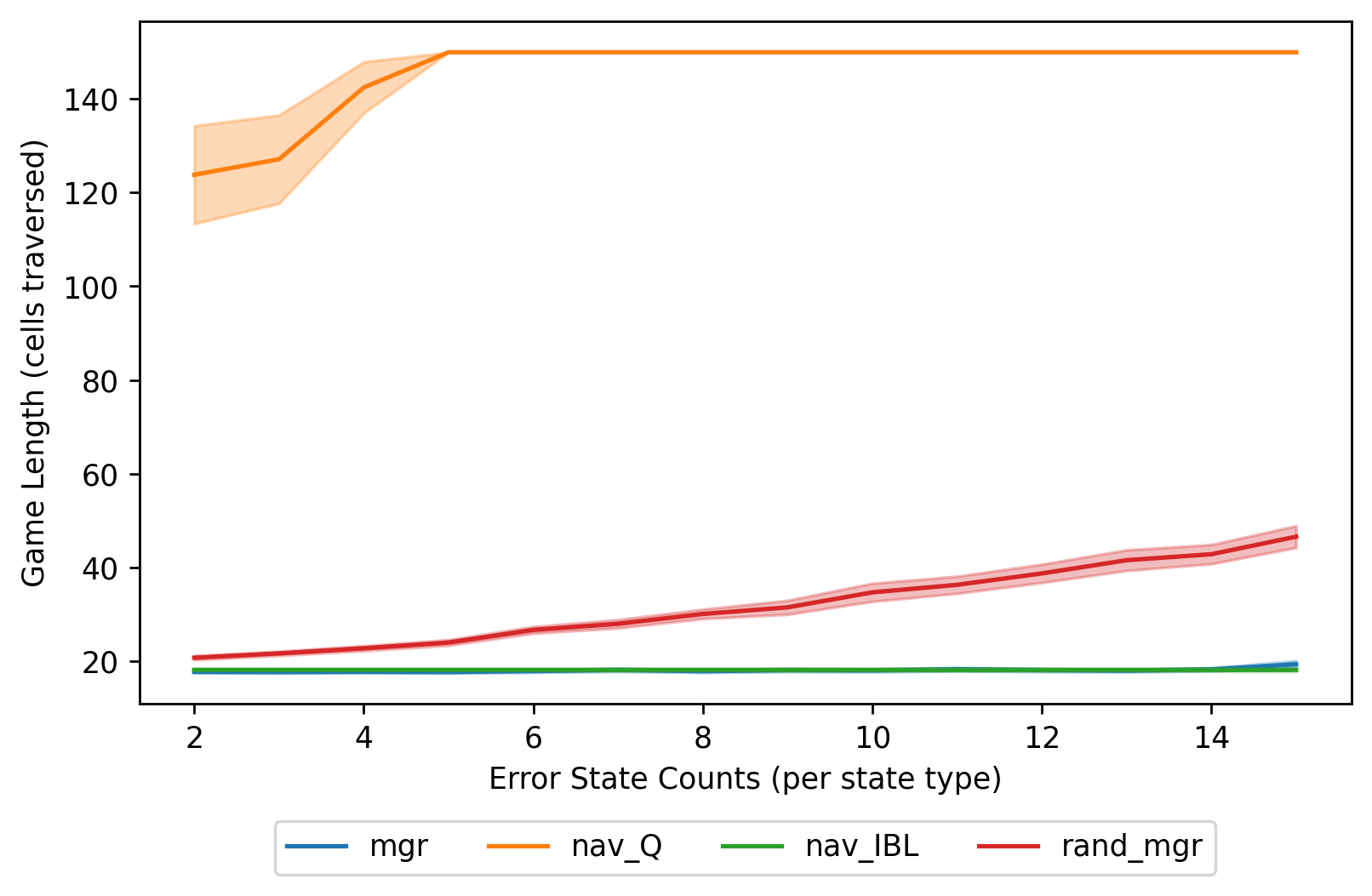}
	\caption{Game Lengths: $\{p_{e_1}=1.0, p_{e_2}=0.0\}$}
	\label{fig:game_lens_60_100_0}
\end{subfigure}
\caption{Game lengths per grid type}
\vspace{-7mm}
\end{figure}

In the first set of results, we tested a team of two navigating agents (one trained with Q-learning and the second with IBL, thus representing an AI agent and a human, respectively) with equal error frequencies in error states for each agents. We hereafter show the results for $p_{e_i} = 0.25$ for each agent $i\in\{1,2\}$, while we omit results obtained with higher error frequencies due to space reasons. The observed behavior is qualitatively equivalent, with performance of solo agents decreasing proportionally as the error frequencies increase. We show in detail results for relatively low error frequencies as, in this case, we expected a diminished effect of errors on game lengths and team performance with respect to the case of higher frequencies. This allows us to show improvements brought by our scheme even in such cases where solo agents already perform well. \autoref{fig:game_lens_60_25_25} shows the game lengths at different error frequencies, for the solo agents (\texttt{nav$_Q$} and \texttt{nav$_{IBL}$} respectively), and for the cases with the random manager and the IBL manager. While the error frequencies are the same, the IBL solo agents perform worse than the Q-learning solo agents because the IBL agent in our case is less likely to perform exploration steps in comparison with the Q-Learning agent. The IBL agent is therefore more likely to develop an earlier bias in its policy and have a larger number of under-explored areas, which would lead to decreased performance when forced off its optimal path. Though, the policy with an IBL manager significantly outperforms a random manager, and is able to improve performance with respect to either solo agents, despite relatively low error frequencies.

\begin{figure}[t]
\centering
\begin{subfigure}{.45\textwidth}
	\centering
	\includegraphics[width=0.9\textwidth]{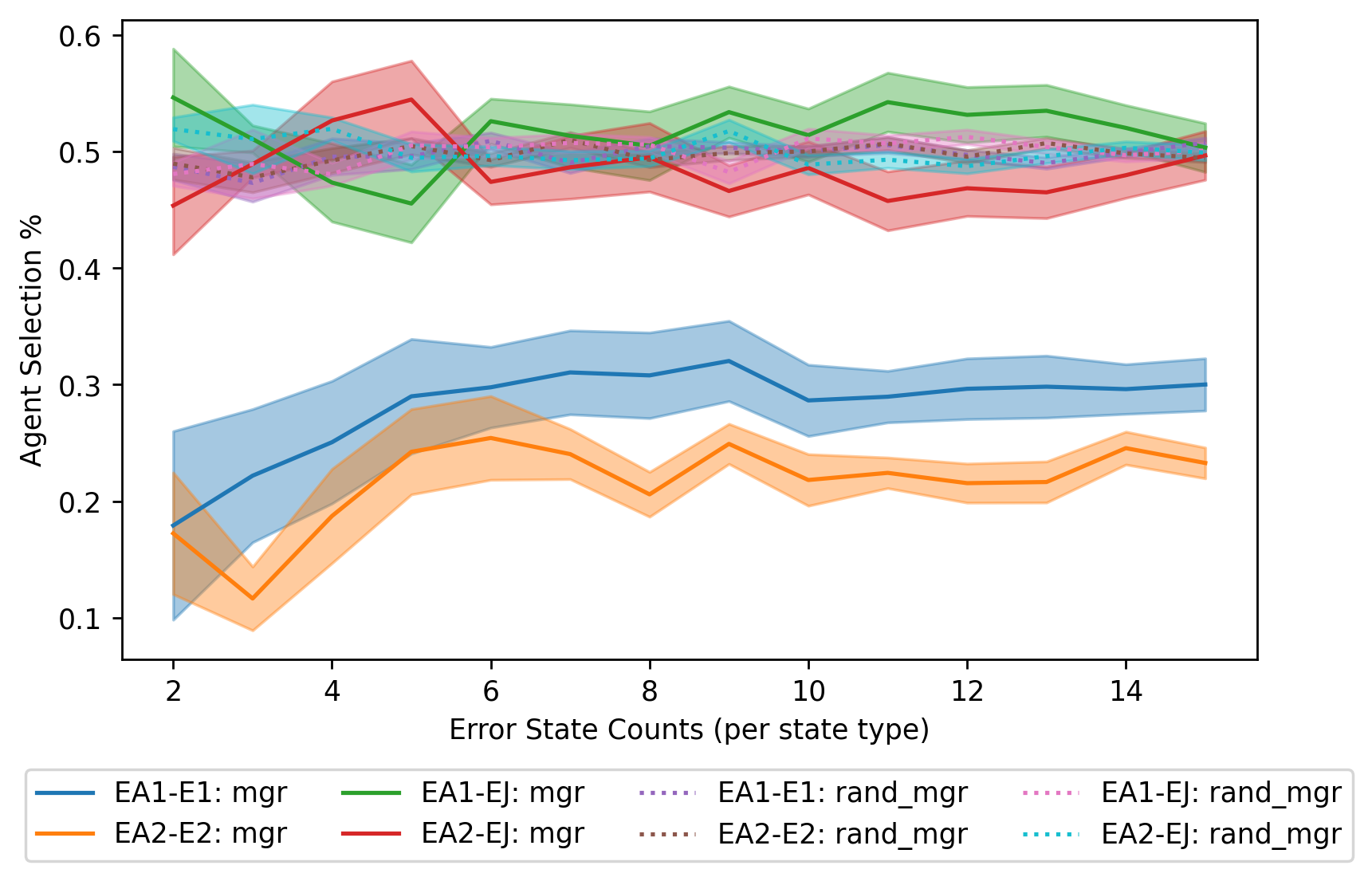}
	\caption{Selection Preferences: $\{p_{e_1}=0.25, p_{e_2}=0.25\}$}
	\label{fig:selection_60_25_25}
\end{subfigure}
\begin{subfigure}{.45\textwidth}
	\centering
	\includegraphics[width=0.9\textwidth]{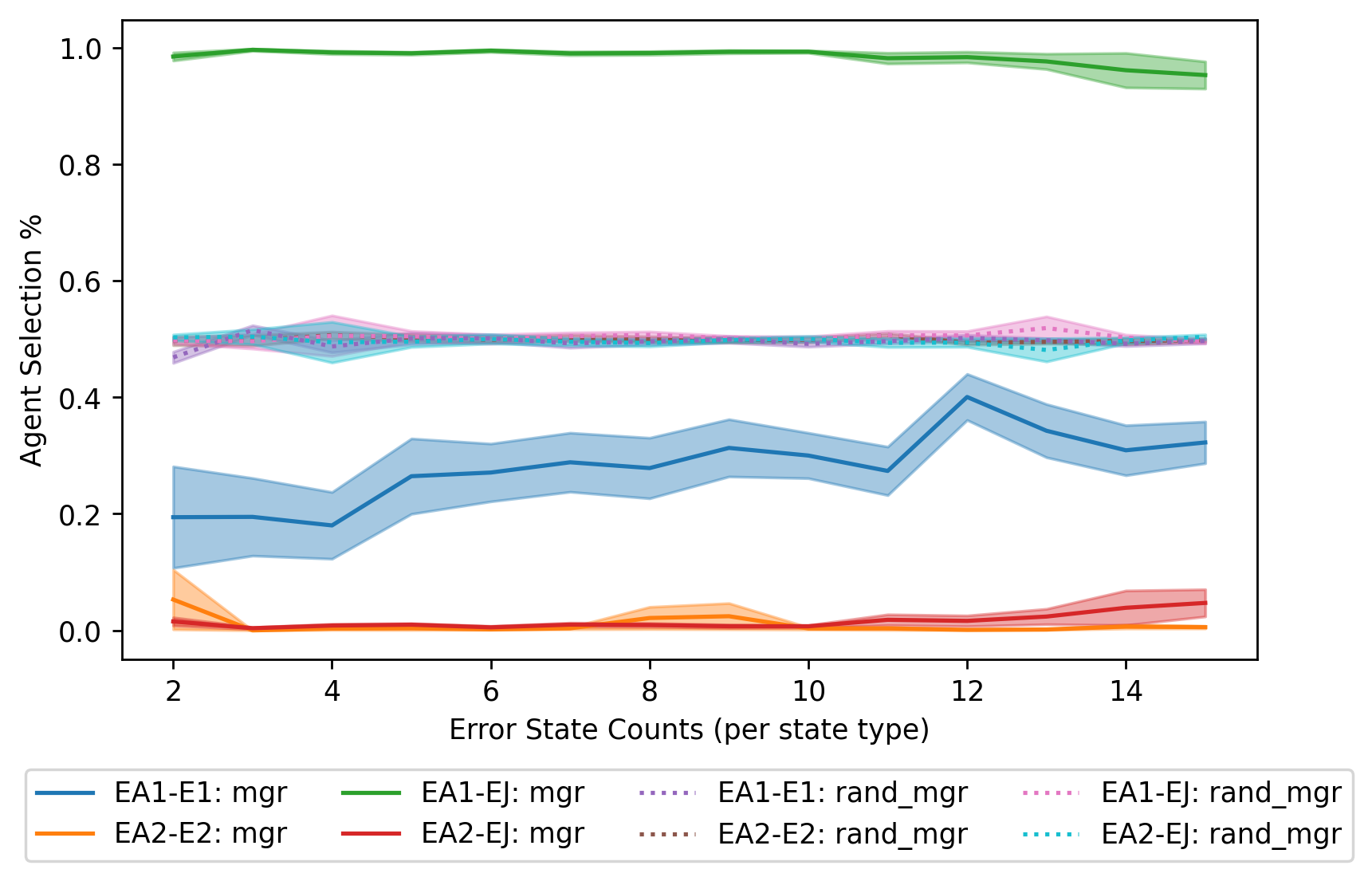}
	\caption{Selection Preferences: $\{p_{e_1}=0.25, p_{e_2}=0.75\}$}
	\label{fig:selection_60_25_75}
\end{subfigure}
\begin{subfigure}{.45\textwidth}
	\centering
	\includegraphics[width=0.9\textwidth]{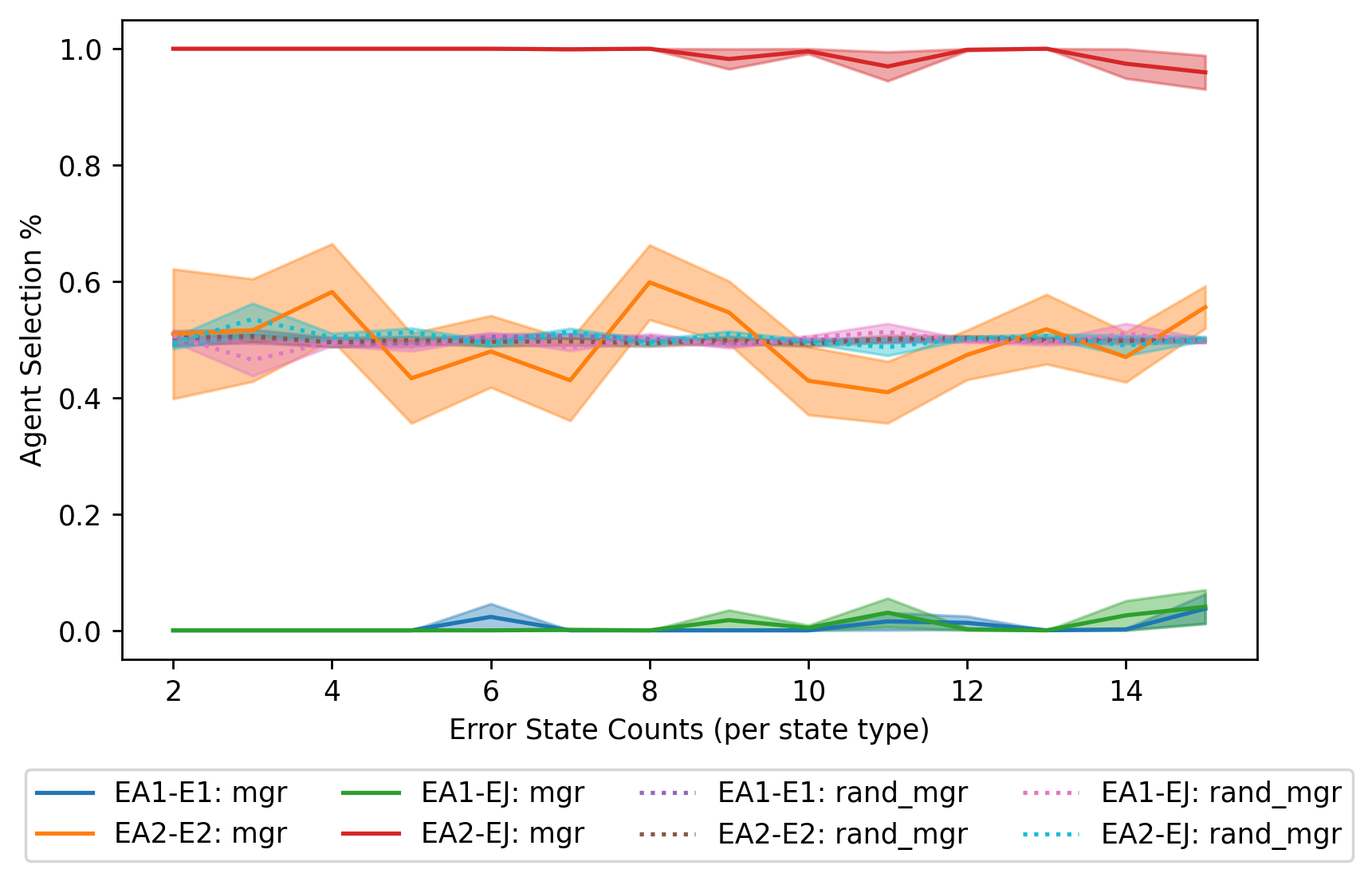}
	\caption{Selection Preferences: $\{p_{e_1}=1.0, p_{e_2}=0.0\}$}
	\label{fig:selection_60_100_0}
\end{subfigure}
\caption{Manager preference for agents per grid type}
\vspace{-7mm}
\end{figure}

In another aspect of performance, the proportion in which the navigating agents demonstrate errors and the IBL manager preference in selecting them show a likely correlation. \autoref{fig:selection_60_25_25} shows the pattern of selection probability. Specifically, curve \texttt{EAx-Ey: mgr} shows the percentage of times the IBL manager selects agent \texttt{x} in error state of type \texttt{y} (\texttt{y=J} for joint error states).

Seen in \autoref{fig:selection_60_25_25}, the manager learns a strong preference for selecting the agents which are error-free for a given error state. On the other hand, the manager demonstrates a lack of complete bias toward the error-free agents. This can likely be attributed to two items. First, the reward signal is based solely on trajectory lengths, so most outcomes (approximately $75\%$) where a low error probability agent is selected will be identical to one with only error-free agent selections. Second, the manager has no prior information on error probabilities, so it's understanding of agent desirability is entirely based on observation and imperfect memory, so the manager learns a preference toward lower error likelihoods without developing a complete bias. The frequency of agent selection maintains this pattern as the frequency of error states of each type is increased through the test cases, which shows awareness of error likelihood by the IBL manager. Notice the difference with respect to the random manager, which picks navigating agents with the same probability of 50\%.

In the next set of results (\autoref{fig:game_lens_60_25_75}, \autoref{fig:selection_60_25_75}), we analyze the behavior with an imbalance in error frequencies for the solo agents, setting error probabilities as $p_{e_1}=0.25, p_{e_2}=0.75$. For this configuration, the impact of agent errors on game lengths is significantly higher for the second agent and so we would expect to see a much higher impact resulting from intelligent selection of agents. The significance of the error likelihood is also expected to impact the successful navigation of the error-prone agent operating in the solo case. The results indicate both of these factors. We can clearly see that the agent with high error likelihood suffers a significant increase in game lengths when operating independently. Additionally, the case of the random manager suffers more in the event of selecting the second agent, resulting in diminished performance. Regarding the IBL manager preferences in agent selection, we see again the preference aligns strongly with error likelihoods. The manager develops a very strong bias toward the better performing agent in joint error states, almost exclusively selecting the low error probability agent. This demonstrates a much stronger belief the second agent will encounter errors from the manager's perspective.

In the final case (\autoref{fig:game_lens_60_100_0}, \autoref{fig:selection_60_100_0}), we demonstrate the impact on performance in the case of a highly divergent team. The error probabilities are $p_{e_1}=1.0, p_{e_2}=0.0$, which demonstrate a worst-case scenario where one agent cannot avoid errors when an error state is encountered. In such a case, we expect the manager to learn a strong bias to the error-free agent in $s_{e_1}$ and $s_{e_{12}}$. On the other hand, in $s_{e_2}$ the manager can treat the two as identical as both have a $0\%$ chance of error. In fact, this pattern is generally demonstrated in \autoref{fig:selection_60_100_0}. We see the manager showing little preference between the two agents in $s_{e_{12}}$. Regarding the solo agent cases, as expected, we also only see the impact on game length for the error-prone agent. Still, we again see the improvement possible with the inclusion of a learning manager. Further, the learning manager shows significantly stronger performance in comparison to the random manager. Additional results are demonstrated in \cite{fuchs2022demo}.

\section{Conclusion}

In this paper, we considered the case of humans and AI systems operating as a team in order to accomplish a task. For our scenario, we tested the combination of a manager agent with a team of agents navigating a Gridworld environment together. We tested a team in which we represented both human and AI behavior through models utilizing standard RL and cognitively inspired IBL models with the inclusion of errors in their behavior. This mixture served to demonstrate a case in which fallible humans and AI systems might operate together to accomplish a shared task. The manager was able to learn a preference for more successful agents and provide a significant improvement in team performance over individual performance and random agent selection. These results show a cognitively inspired model can be utilized to learn from patterns of behavior to effectively coordinate behavior in the event the team shows signs of producing errors.

\bibliographystyle{IEEEtran}
\bibliography{IEEEabrv,main}

\end{document}